\documentclass{PoS}

\title{Recent Results of Electron-Neutrino Appearance Measurement at NOvA}

\ShortTitle{Recent Results of Electron-Neutrino Appearance Measurement at NOvA}

\author{\speaker{Jianming Bian (for the NOvA Collaboration)}\thanks{}\\
        University of California, Irvine\\
        E-mail: \email{bianjm@uci.edu}}

\abstract{NOvA is a long-baseline accelerator-based neutrino oscillation experiment that is optimized for $\nu_e$ measurements. It uses the upgraded NuMI beam from Fermilab and measures electron-neutrino appearance and muon-neutrino disappearance at its Far Detector in Ash River, Minnesota. The $\nu_e$ appearance analysis at NOvA aims to resolve the neutrino mass hierarchy problem and to constrain the CP-violating phase. The first measurement of electron-neutrino appearance in NOvA based on its first year's data was produced in 2015, providing solid evidence of $\nu_e$ oscillation with the NuMI beam line and some hints on mass-hierarchy and CP. This talk will discuss the second $\nu_e$ oscillation analysis at NOvA, which is based on 2 years of data.}

\FullConference{38th International Conference on High Energy Physics\\
		3-10 August 2016\\
		Chicago, USA}

\begin{document}

\section{Introduction}
NOvA is a long-baseline neutrino experiment optimized to observe the oscillation of muon neutrinos to electron-neutrinos. NOvA uses a 14-kt liquid scintillator Far Detector (FD) in Ash River, Minnesota to detect the oscillated NuMI (Neutrinos at the Main Injector) muon neutrino beam produced 810 km away at Fermilab. The NuMI beam has been upgraded to  700 kW. NOvA has the longest baseline in operation, so the matter effect is as large as $30\%$, which is sensitive to the mass hierarchy determination.  NOvA is equipped with a  0.3-kt functionally identical Near Detector (ND) located at Fermilab to measure unoscillated beam neutrinos and estimate backgrounds at the FD. Both detectors are located 14.6 mrad off-axis to receive a narrow-band neutrino energy spectrum near the energy of the $\nu_\mu\to\nu_e$ oscillation maximum range ($\sim$2 GeV), enhancing the $\nu_\mu\to\nu_e$ oscillation signal in the FD while reducing neutral current and beam $\nu_e$ backgrounds from high-energy unoscillated beam neutrinos. 

NOvA detectors are composed of PVC modules extruded to form long tube-like cells. The  FD consists of 344,064 detector cells, and the ND consists of 20,192 cells. Lengths of cells are 16m in FD and 4m in ND. Each cell is filled with liquid scintillator and has a loop of wavelength-shifting fiber routed to an Avalanche Photodiode. The cells are arranged in planes, assembled in alternating vertical and horizontal directions, so the NOvA detectors can serve as both calorimeters and 3-D trackers. The NOvA detectors have low-Z and low-density, each plane is just 0.15 $X_0$, which is great for $e/\pi^0$ separation.

The $\nu_e$ appearance analysis at NOvA aims to determine the neutrino mass hierarchy, CP violation and the octant of $\theta_{23}$.  Using the first year of neutrino beam data ($2.74\times10^{20}$ POT), the NOvA collaboration has published first papers that present its initial results of the $\nu_e$ appearance and $\nu_\mu$ disappearance measurements~\cite{Adamson:2016tbq}\cite{Adamson:2016xxw}. For the analysis reported in this paper, we use $6.05\times10^{20}$ POT neutrino beam data, which is more than twice as large as the exposure in our first analysis last year.

\section{Event selection and Data Analysis}

Recently, a convolutional neural network based algorithm has been implemented at NOvA to serve as the primary event identifier for the second analysis. The new particle identification (PID) algorithm is named CVN~\cite{Aurisano:2016jvx}.  It uses pixels as inputs and the output is a variable that describes the probability to be a $\nu_e$ CC event with a range $0-1$. In this CVN identifier, convolutional filters are used to automatically extract features from the raw hit map. The output of this neural net is used to classify the event. The statistical power of CVN is equivalent to $30\%$ more exposure than previous PIDs, LID~\cite{Bian:2015opa} and LEM~\cite{Backhouse:2015xva}. In this analysis, an empirical meson exchange current (MEC) model~\cite{Rodrigues:2015hik} is added in the GENIE event generator~\cite{Andreopoulos:2009rq}.  According to the hadronic energy distributions in the ND $\nu_\mu$ CC Data/MC, this process was found missing in our default simulation in the phase space between the quasi-elastic (QE) and the resonance production  (RES) regions, which is also reported by MINERvA~\cite{Rodrigues:2015hik}. At NOvA, these simulated MEC events are reweighted based on Data/MC comparisons of kinetic variables of ND $\nu_\mu$ CC events.

\begin{figure}[h]
\centering
\includegraphics[width=12cm]{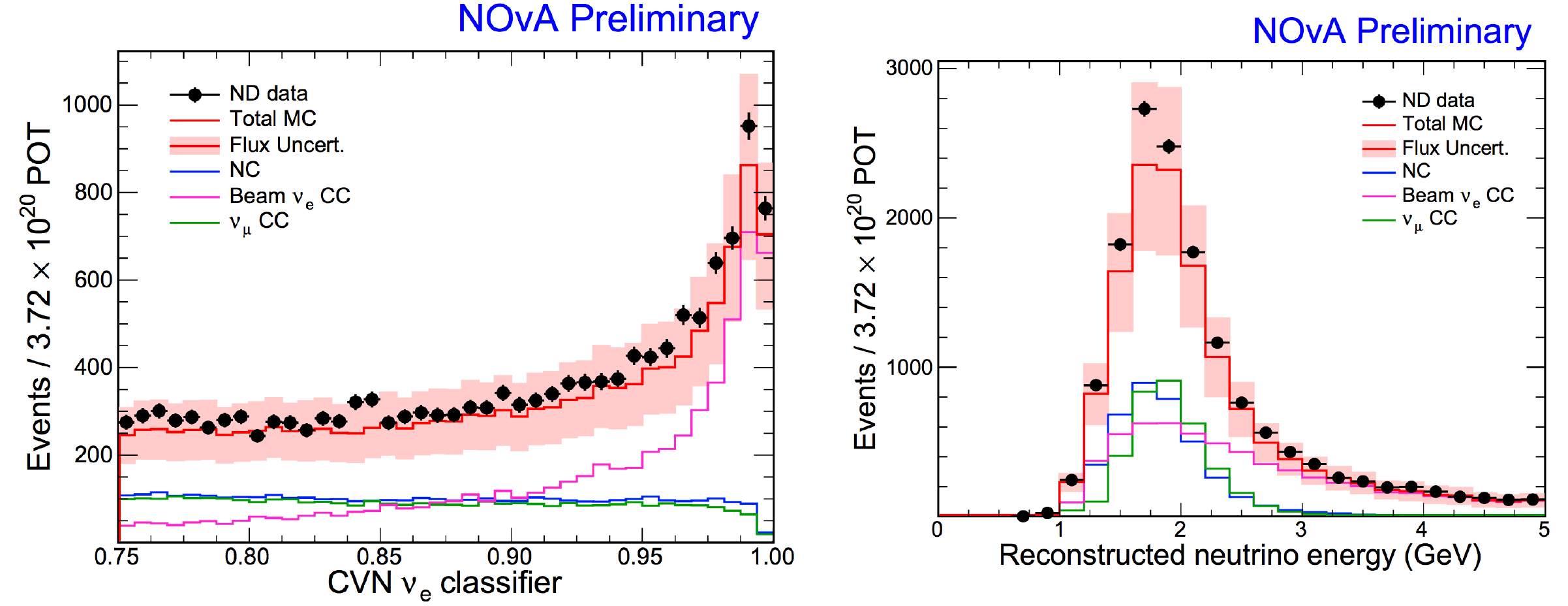}
\caption{Data and MC of the CVN distribution (left) and the reconstructed neutrino energy (right) in the near detector}\label{fig:nddatamc}
\end{figure}

We optimize event selection to maximize $FOM=S/\sqrt{S+B}$. According to the far detector MC, the signal efficiency after selection is $73\%$, and the purity is $76\%$. After the event selection, we reconstruct the neutrino energy as a function of the electromagnetic energy and hadronic energy. Data and MC of the CVN distribution and the reconstructed neutrino energy in the near detector are shown in Figure~\ref{fig:nddatamc}.  We then use the ND data to predict the background energy spectrum in the far detector in 3 PID regions.

For the $\nu_e$ CC signal, the consistency between data and MC has been validated based on the ND data and the cosmic ray data, using beam $\nu_\mu$ CC (replace the muon track with a simulated electron shower) events and electromagnetic showers in cosmic rays. Signal Data/MC difference is found to be less than $1\%$. Beam $\nu_e$, $\nu_\mu$ and NC backgrounds are extrapolated differently to the FD. We first reweight kaon and pion components in the flux to match the selected $\nu_{\mu}$CC energy spectrum in data, then fix the beam $\nu_e$ to the flux-reweighted result (up by $4\%$), and constrain NC (up by $10\%$) and $\nu_\mu$ CC (up by $17\%$) using Michel electron distributions. After this tuning,  the simulated energy spectrum matches the data in all three PID bins.

\begin{figure}[h]
\centering
\includegraphics[width=8cm]{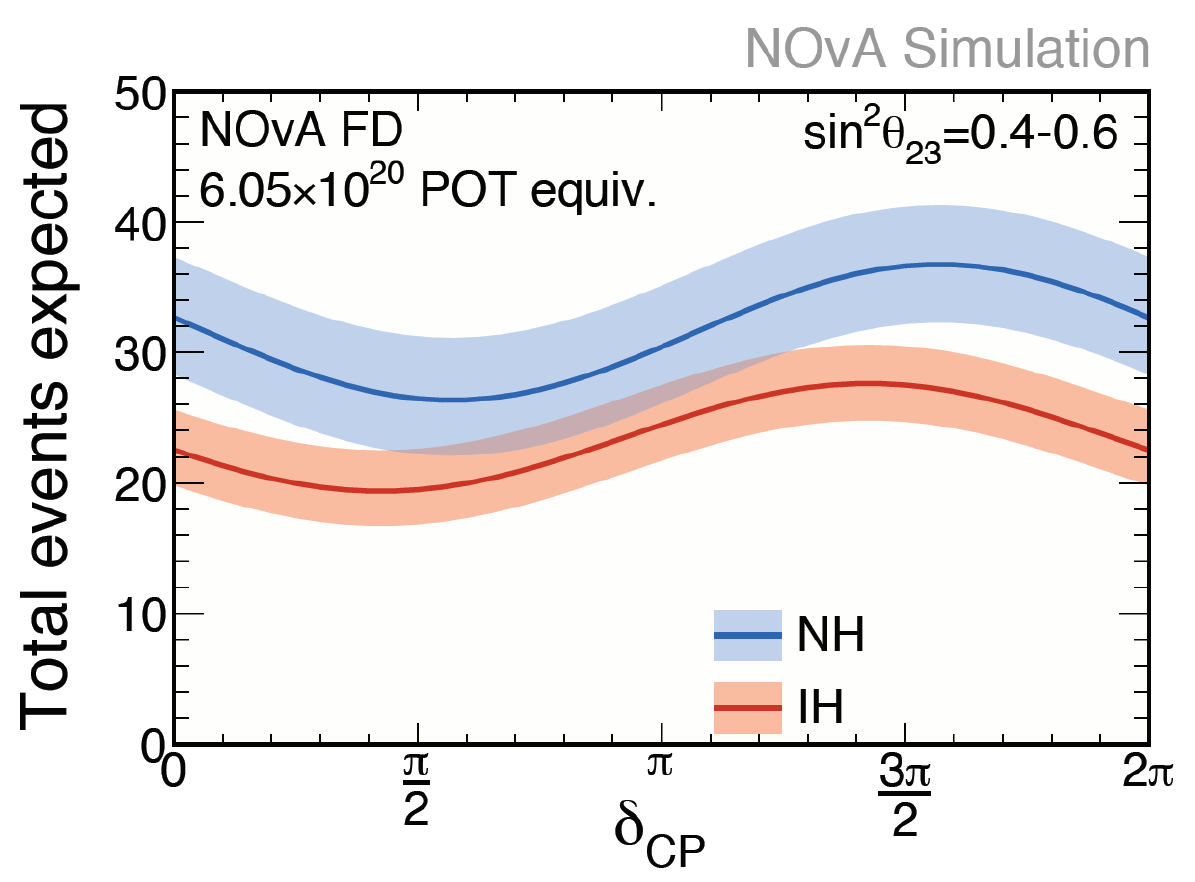}
\caption{Total number expected event vs. CP phase}\label{fig:evtvscp}
\end{figure}

We predict FD signal+background counts based on the signal MC, the ND data and the cosmic ray data (FD events outside of the beam spill window). The total expected event count depends on oscillation parameters, as shown in Figure~\ref{fig:evtvscp}. Under the assumption of normal mass hierarchy, $\delta_{CP}=3/2\pi$ and $\sin^2 \theta_{23}=0.5$, we expect  to have 36.4 events in total with the CVN selection. When assuming inverted mass hierarchy and $\delta_{CP}=1/2\pi$, the expected number of events is 19.4.  The background prediction in the FD is 8.2 events, varying about $\%1$ for different choices of oscillation parameters. The dominant backgrounds are 3.7 NC events and 3.1 beam $\nu_e$ CC events. $\nu_\mu$ CC, $\nu_\tau$ CC and cosmic muons take 0.7, 0.1 and 0.5 events, respectively.


This extrapolation eliminates most of the systematic errors. Remaining systematic uncertainties after the extrapolation are evaluated by extrapolating ND data with nominal MC and systematically modified MC samples, with variations to the normalization (POT counting), neutrino cross-sections, calibration, beam simulation, non-linearity in detector responses and other smaller uncertainties. The total systematic errors are about $5\%$ for the signal and about $10\%$ for the background.

\section{Results}

Before counting and analyzing the FD data in the beam spill window (the signal region), we checked the near-PID sideband, the high-energy sideband and events outside of the beam spill window. They all appeared to have good data/MC agreements.  Using the CVN selection, we observe 33 $\nu_e$ candidates in the FD, and the expected background is $8.2\pm0.8$ events. Using LID and LEM, event selectors in the 2015 analysis, we observe consistent results (LID: 34 events, $12.2\pm1.2$ background expected; LEM: 33 events, $10.3\pm1.0$ background expected).

\begin{figure}[h]
\centering
\includegraphics[width=12cm]{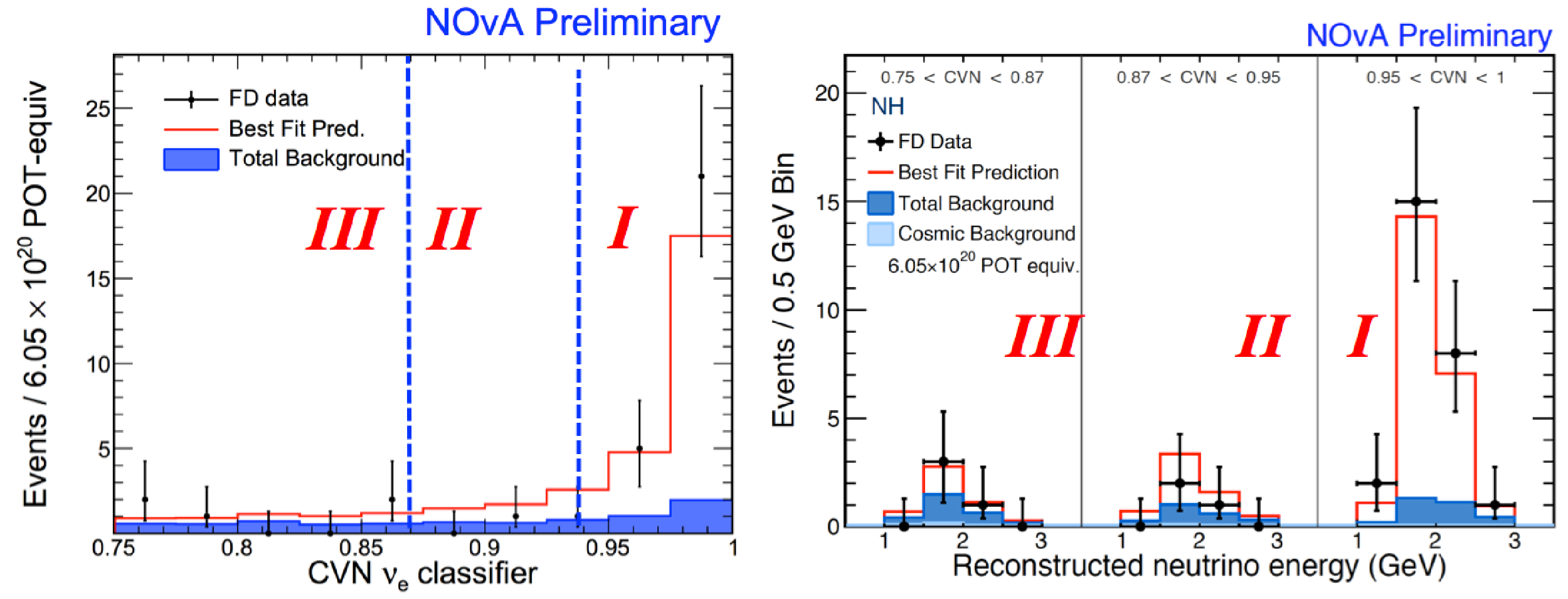}
\caption{The CVN distribution (left) and the reconstructed energy spectrum (right) of $\nu_e$ CC selected events in the far detector. Black points: FD data, red histogram: the best fit prediction, blue shaded histograms: background MC.}\label{fig:fddatamc}
\end{figure}

Energy distributions in the three CVN bins are fit to data to extract oscillation parameters. In this fit, oscillation parameters are constrained as: $\sin^22\theta_{13}=0.086\pm0.05$, $\Delta m^2_{21}=7.53\pm 0.18 \times 10^{-5}$eV$^2$ and $\Delta m^2_{32}=2.44\pm0.06\times10^{-3}$ eV$^2$ for the normal mass hierarchy (NH) and $-2.49\pm0.06\times10^{-3}$ eV$^2$ for the inverted mass hierarchy (IH). The systemic uncertainties are included as nuisance parameters in the fit. Figure~\ref{fig:fddatamc} shows comparisons of the FD data to the best fit prediction for CVN and energy distributions, under the normal hierarchy assumption. The significance of $\nu_e$ appearance is greater than $8 \sigma$. 

\begin{figure}[t]
\centering
\includegraphics[width=12cm]{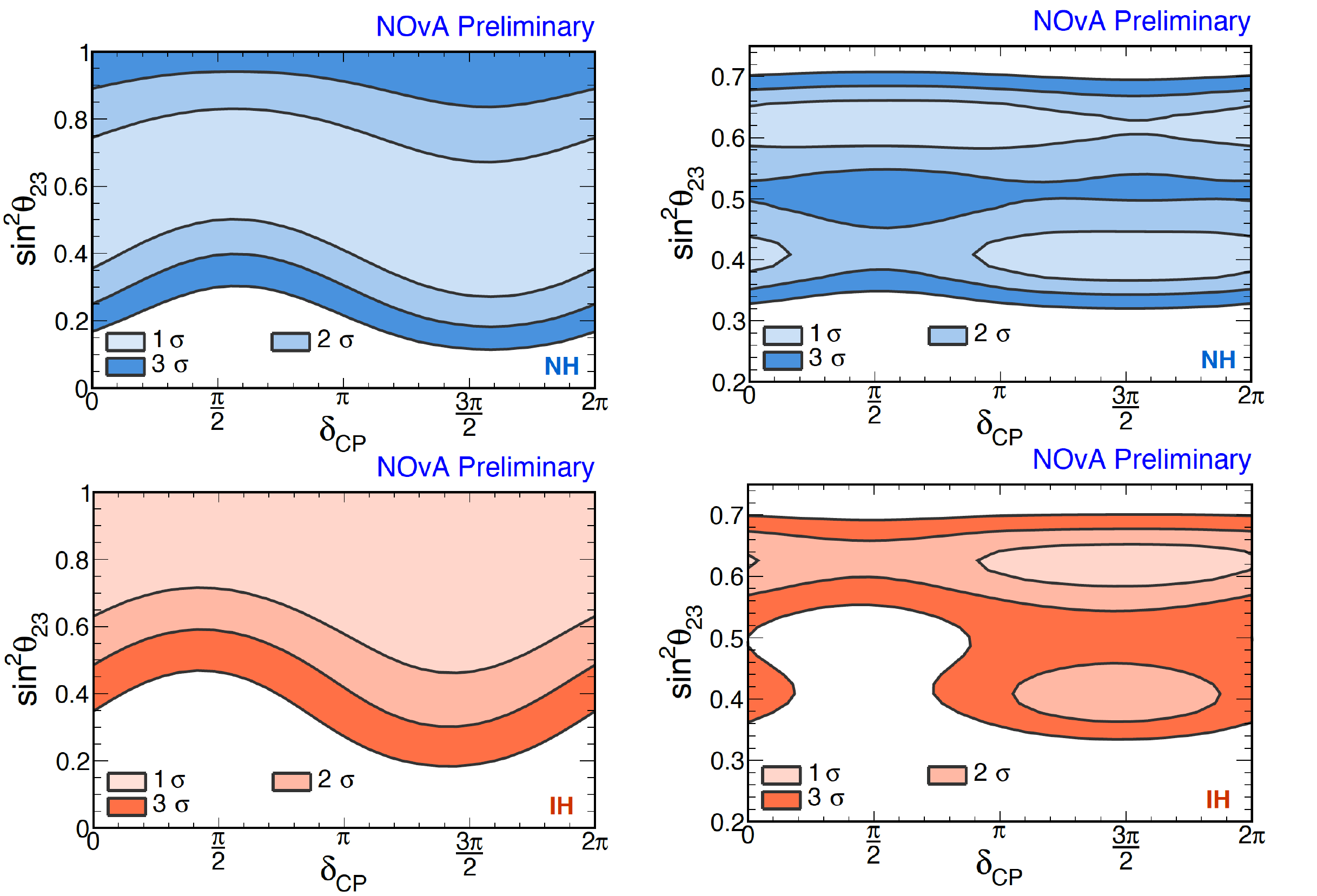}
\caption{Allowed regions of $\sin^2\theta_{23}$ vs. $\delta_{CP}$ at 1, 2, and $3\sigma$: (left) results from $\nu_e$ appearance data and (right) results from the combination of $\nu_e$ appearance and $\nu_\mu$ disappearance data.}\label{fig:contour}
\end{figure}

The resulting allowed regions of $\sin^2\theta_{23}$ vs. $\delta_{CP}$ at 1, 2, and $3\sigma$ for each of the hierarchies produced by the fit are shown in Figure~\ref{fig:contour}. The left two plots show contours from the $\nu_e$ appearance data and right two plots show contours from the combination of $\nu_e$ appearance and $\nu_\mu$ disappearance data. The global best fit gives us normal hierarchy, $3\pi/2$ CP phase and $\sin^2\theta_{23} = 0.4$.
Because the NOvA $\nu_\mu$ disappearance analysis found that the mixing is not maximal, both octants are allowed for $\theta_{23}$. If it is the lower octant, NH is preferred and CP is close $3\pi/2$. For the upper octant, both MHs are allowed, and the best fit is around $\pi/2$ for NH, and $3\pi/2$ for IH. $\delta_{CP} \simeq \pi/2$ is rejected for the IH and lower octant.

\begin{figure}[h]
\centering
\includegraphics[width=8cm]{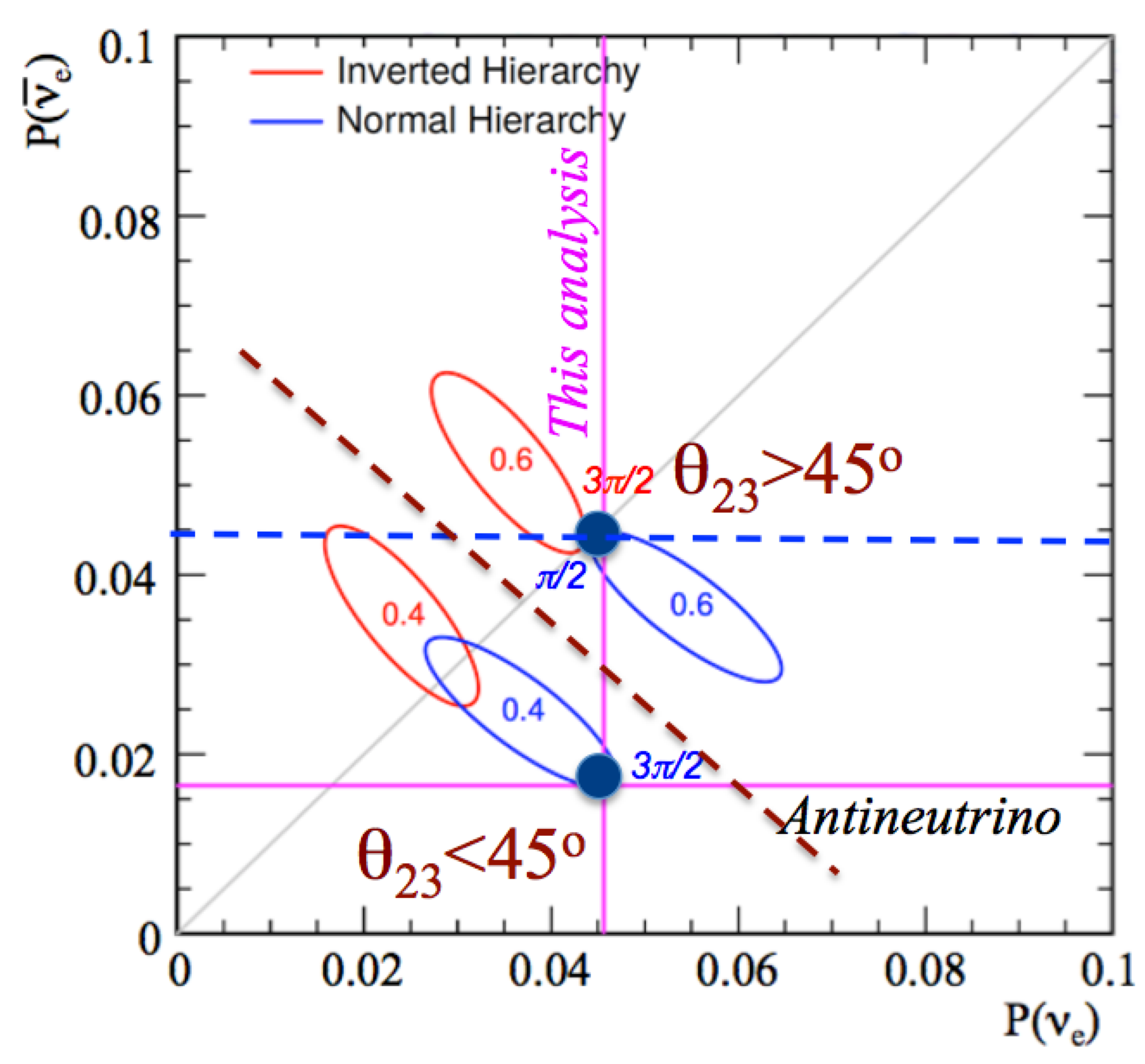}
\caption{The principle of the NOvA $\nu_e(\bar{\nu}_e)$ appearance measurements. All possible values of probabilities of $\nu_\mu\to\nu_e$ vs. $\bar{\nu}_\mu\to\bar{\nu}_e$ are on the ellipses. The solid blue (red) ellipses correspond to the normal (inverse) hierarchy scenarios, with $\delta_{CP}$ varying as one moves around each ellipse.}\label{fig:biplot}
\end{figure}

Figure~\ref{fig:biplot} demonstrates the ambiguity caused by the octant of $\theta_{23}$ in our results. The ellipses are all possible results of the $\nu_e$ appearances vs. $\bar{\nu}_e$ appearance probabilities in NOvA. For $\theta_{23}<45^\circ$, the NH and IH ellipses are on the lower left side. For $\theta_{23}>45^\circ$, the two ellipses are on the upper right side. The result of this analysis ($\nu_\mu\to\nu_e$) is the vertical violet line.  One can find that our results prefer (NH, $3\pi/2$ CP) for the lower octant, and prefer (IH, $3\pi/2$) or (NH, $\pi/2$) for the upper octant. This ambiguity can be solved by performing a measurement of $\bar{\nu}_\mu\to\bar{\nu}_e$ (horizontal violet line) using the anti-neutrino beam.

\section{Summary}

With $6.05\times10^{20}$ of POT NuMI neutrino beam data, we have performed the second $\nu_e$ appearance measurement in NOvA. The significance of the $\nu_e$ appearance is greater than $8\sigma$, and our data prefers the normal mass hierarchy at a low significance. In our analysis, the inverted mass hierarchy for $\delta_{CP}=\pi/2$ is rejected for lower octant, but we do observe an ambiguity in the mass hierarchy determination caused by the octant of $\theta_{23}$. We plan to run antineutrinos to perform a $\bar{\nu}_e$ appearance analysis in Spring, 2017 to solve these degeneracies.

\end{document}